\documentclass[a4paper, 10 pt, conference]{ieeeconf}  

\IEEEoverridecommandlockouts                              

\usepackage{graphicx}
\usepackage[latin1]{inputenc}

\usepackage{amssymb,amsmath,amsfonts,mathtools}

\usepackage[dvipsnames]{xcolor}
\usepackage[noadjust]{cite}

\usepackage{amsthm}
\usepackage{tikz}

\usepackage{algorithm}
\usepackage{algpseudocode}
\usepackage{algorithmicx}
\usepackage{booktabs,tabularx}

\usepackage[official]{eurosym}

\makeatletter
\let\NAT@parse\undefined
\makeatother
\usepackage{hyperref}
\hypersetup{
	colorlinks=true,
	linkcolor=Blue,
	urlcolor=Blue,
	citecolor=Blue,
}

\newcommand{\cE}{\mathcal{E}}
\newcommand{\cG}{\mathcal{G}}
\newcommand{\cW}{\mathcal{W}}

\newcommand{\cD}{\mathcal{D}}
\newcommand{\cN}{\mathcal{N}}

\newcommand{\mc}{\mathcal}

\newcommand{\R}{\mathbb{R}}
\newcommand{\N}{\mathbb{N}}

\newcommand{\x}{x}

\newcommand{\w}{w}
\newcommand{\z}{z}
\newcommand{\y}{y}
\newcommand{\X}{X}

\newcommand{\iter}{t}
\newcommand{\iterp}{t+1}
\newcommand{\xtp}{\x^{\iterp}}

\newcommand{\wtp}{\w^{\iterp}}
\newcommand{\ztp}{\z^{\iterp}}

\newcommand{\xt}{\x^{\iter}}

\newcommand{\wt}{\w^{\iter}}
\newcommand{\zt}{\z^{\iter}}

\newcommand{\Wd}{\cW_d}

\newcommand{\xstar}{\x^\star}
\newcommand{\xistar}{\x^\star_i}
\newcommand{\xstari}{\x^\star_{-i}}

\newcommand{\xitp}{\x_{i}^{\iterp}}

\newcommand{\zitp}{\z_{i}^{\iterp}}

\newcommand{\xit}{\x_{i}^{\iter}}

\newcommand{\zit}{\z_{i}^{\iter}}

\newcommand{\xjt}{\x_{j}^{\iter}}

\newcommand{\zjt}{\z_{j}^{\iter}}

\newcommand{\tz}{\tilde{\z}}

\newcommand{\tztp}{\tz^{\iterp}}

\newcommand{\psitp}{\psi^{\iterp}}

\newcommand{\tzt}{\tz^{\iter}}

\newcommand{\psit}{\psi^{\iter}}

\newcommand{\agg}{\sigma}
\newcommand{\bz}{\bar{\z}}

\newcommand{\pz}{\z_\perp}

\newcommand{\bzt}{\bz^{\iter}}
\newcommand{\bztp}{\bz^{\iterp}}

\newcommand{\pzt}{\pz^{\iter}}
\newcommand{\pztp}{\pz^{\iterp}}

\newcommand{\xmi}{\x_{-i}}

\newcommand{\sigmai}{\sigma_{-i}}

\newcommand{\chit}{\chi^{\iter}}
\newcommand{\chitp}{\chi^{\iterp}}

\newcommand{\Ji}{J_i}

\newcommand{\F}{F}

\newcommand{\tFi}{\tilde{F}_i}

\newcommand{\tF}{\tilde{F}}

\newcommand{\phii}{\phi_i}
\newcommand{\phij}{\phi_j}

\newcommand{\oned}{\mathbf{1}_{N,d}}

\newcommand{\cC}{\mathcal{C}}

\newcommand{\norm}[1]{\left \|#1 \right \|}
\newcommand{\Px}[1]{P_{\X}\left[#1 \right]}
\newcommand{\Pxi}[1]{P_{\X_i}\left[#1 \right]}

\newcommand{\Rd}{R_{d}}

\newcommand{\lwm}{M} 

\newcommand{\aggmi}{\agg_{-i}}

\newcommand{\T}{^\top}
\newcommand{\map}[3]{#1: #2 \rightarrow #3}

\newcommand{\col}{\mathrm{col}}

\newcommand\oprocendsymbol{\hbox{$\square$}}
\newcommand\oprocend{\relax\ifmmode\else\unskip\hfill\fi\oprocendsymbol}

\def\er/{Erd\H{o}s-R\'enyi}

\graphicspath{{figs/}}

\def\algo/{{TRADES}}
\def\algoc/{{Primal-Dual TRADES}}

\newtheorem{theorem}{Theorem}[section]

\newtheorem{definition}[theorem]{Definition} 
\newtheorem{lemma}[theorem]{Lemma}

\newtheorem{assumption}[theorem]{Assumption}

\begin{document}
\title{Distributed equilibrium seeking in aggregative games:\\ linear convergence under singular perturbations lens}
\author{Guido Carnevale, Filippo Fabiani, Filiberto Fele, Kostas Margellos, Giuseppe Notarstefano
\thanks{G.~Carnevale and G.~Notarstefano are with the Department of Electrical,  Electronic and Information Engineering,  Alma Mater Studiorum - University of Bologna, Italy (\texttt{guido.carnevale@unibo.it; giuseppe.notarstefano@unibo.it}); F.~Fabiani is with the IMT School for Advanced Studies Lucca, Italy ({\tt filippo.fabiani@imtlucca.it}); F.~Fele is with the Department of Systems Engineering and Automation, University of Sevilla, Spain (\texttt{ffele@us.es}); K.~Margellos is with the Department of Engineering Science, University of Oxford, UK (\texttt{kostas.margellos@eng.ox.ac.uk}).
}
\thanks{\copyright 2024 IEEE. Personal use of this material is permitted.  Permission from IEEE must be obtained for all other uses, in any current or future media, including reprinting/republishing this material for advertising or promotional purposes, creating new collective works, for resale or redistribution to servers or lists, or reuse of any copyrighted component of this work in other works. Final version of the article available at \url{https://doi.org/10.1109/CDC56724.2024.10886119} (please cite as \cite{GuidoEtAl2024CDC}).}
}

\maketitle

\begin{abstract}
	We present a fully-distributed algorithm for Nash equilibrium seeking in aggregative games over networks. 
    The proposed scheme endows each agent with a gradient-based scheme equipped with a tracking mechanism to locally reconstruct the aggregative variable, which is not available to the agents.
    We show that our method falls into the framework of \emph{singularly perturbed systems}, as it involves the interconnection between a fast subsystem -- the global information reconstruction dynamics -- with a slow one concerning the optimization of the local strategies.
    This perspective plays a key role in analyzing the scheme with a constant stepsize, and in proving its linear convergence to the Nash equilibrium in strongly monotone games with local constraints.
    By exploiting the flexibility of our aggregative variable definition (not necessarily the arithmetic average of the agents' strategy), we show the efficacy of our algorithm on a realistic voltage support case study for the smart grid.
\end{abstract}

\section{Introduction}
\label{sec:introduction}

The computation of Nash equilibria (NE) in games over networks~\cite{facchinei2007gnep} is receiving increasing attention due to their wide applicability, for instance, in smart grids~\cite{deori2018price,fele2020probably,cenedese2019charging} and power grids~\cite{yin2011synchronization}, or cooperative and network control \cite{fabiani2018distributed}.

\emph{Aggregative} games \cite{Jensen2010,FeleEtAl2018aggrRH,belgioioso2021semi} constitute a prominent class, exhibiting a specific structure often encountered in practice, where the cost of each agent in the network is influenced by its competitors via the so-called aggregative variable.
In this paper, we develop a procedure in which each agent iteratively shares their tentative strategy, subject to local constraints, only with some agents in the network -- namely, neighbors according to a given communication graph -- ultimately leading to an aggregative NE.
Although the aggregative variable depends on the strategies of all the agents in the network and affects each local function,  our distributed strategy does not require its local knowledge.
In particular, our scheme embeds a consensus-based mechanism that locally compensates for this lack of knowledge via neighbors' information exchanges.

Along this direction, available approaches involve semi-decentralized algorithms, where a central authority is assumed to be in place, collecting/broadcasting information from/to all agents in the network~\cite{belgioioso2017semi,grammatico2017dynamic,paccagnan2018nash,yi2019operator,belgioioso2021semi}.
To relax the communication requirements,~\cite{koshal2016distributed} introduces a dynamic averaging consensus methodology based on a gradient-based algorithm with diminishing stepsize (see, e.g.,~\cite{zhu2010discrete}) to reach an agreement on the aggregative variable.
In~\cite{ye2021differentially} this method is further extended to deal with privacy issues, whose convergence is however established for \emph{approximate} NE only.
A fully-distributed method for exact NE computation, instead, is proposed in~\cite{parise2020distributed} leveraging a best-response iterative scheme. This algorithm, however, requires multiple communication rounds per iteration. An asynchronous distributed methodology based on proximal dynamics is provided in \cite{cenedese2020asynchronous}.
Finally, in~\cite{bianchi2020fully,bianchi2022fast} distributed NE seeking algorithms based on proximal best-response are proposed, achieving linear convergence rate (see \cite[Rem.~12]{bianchi2022fast}).

Focusing on aggregative games with local constraints, in this paper we provide a novel, fully-distributed algorithm for NE seeking that exhibits a linear convergence rate. 
Our algorithm capitalizes on recent developments in distributed aggregative optimization~\cite{li2021distributed,li2021distributedOnline,carnevale2022distributed}, and combines an approximated projected pseudo-gradient method with a consensus-based mechanism that asymptotically compensates for the local lack of knowledge of the aggregative variable.  
To establish linear convergence to the NE, we analyze the algorithm dynamics through the lens of \emph{singular perturbation} theory.
As customary in this kind of approach, we identify a \emph{slow} subsystem -- produced by the update of the agents' strategies -- and a \emph{fast} one capturing the consensus-based part of the scheme.
We then separately study two corresponding auxiliary systems related to the identified subsystem to conclude about the stability and convergence properties of the original interconnected scheme.
Since our proposal builds upon a gradient-based scheme, it affords a low computational complexity with respect to comparable best-response algorithms, as the one in \cite{bianchi2022fast}.
Our approach further differs from available ones as it allows for a more general form of aggregative variable -- instead of the standard arithmetic average of the agents' strategies. We exemplify the relevance of this property in a real-world application in \S\ref{subsec:motivating_example}. 
Our theoretical results are finally accompanied by numerical simulations on a voltage control problem for smart grids, thus illustrating the effectiveness of our scheme as well as the flexibility and relevance of the considered setup.

\textit{Notation}:
$I_m$, $0_m \in \R^{m \times m}$, denote the $m\times m$ identity and all-zero matrices, respectively. 
$1_N$ is the vector of $N$ ones, while
$\oned \coloneqq 1_N \otimes I_d$, with $\otimes$ being the Kronecker product (dimensions are omitted when clear from the context). 
Given $f: \R^{n_1} \times \R^{n_2} \to \R$, $\nabla_1 f \in \R^{n_1}$ and $\nabla_2 f \in \R^{n_2}$ denote the gradient of $f$ w.r.t.~its first and second argument, respectively.
For vectors $v_1, \dots, v_N \in \R^n$, $\col (v_1, \dots, v_N)$ denotes their vertical stack. 
For a closed, convex set $\X \subseteq \R^n$ and $x\in \R^{n}$, $\Px{x}$ is the projection of $x$ onto $\X$.

\section{Problem setup and case study}

    \subsection{Problem Setup}

Consider $N \in \N$ agents indexed by the set $\mc I \coloneqq \{1,\dots, N\}$, whose individual strategy is denoted as $\x_i \in \R^{n_i}$. Given all other players' strategies, each agent aims at solving the following optimization problem:
\begin{equation}\label{eq:problem}
	\forall i \in \mc I : \underset{\x_i\in \X_i}{\min} \; \Ji(\x_i, \agg(\x)),
\end{equation}
where $x \coloneqq \col(\x_1,\dots,\x_N) \in \R^n$ is the global strategy vector, and the cost function $\map{\Ji}{\R^{n_i} \times \R^d}{\R}$ depends on $\x_i \in \R^{n_i}$ as well as on the \emph{generalized aggregator} \cite{Jensen2010} 
\begin{align}\label{eq:sigma}
	\agg(\x) \coloneqq \frac{1}{N} \sum_{i \in \mc I}\phii(\x_i),
\end{align}
where each \emph{aggregation rule} $\phii: \R^{n_i} \to \R^d$ models the contribution of $\x_i$ to $\sigma(\x)$.
Finally, $\X_i \subseteq \R^{n_i}$ denotes the domain of feasible strategies for agent $i$ and $\X := \prod_{i\in\mc I}\X_i$ is the collective one.
A standard solution concept for the aggregative game in \eqref{eq:problem} is the NE~\cite{facchinei2007gnep} defined as follows.
\begin{definition}[\textup{Nash equilibrium}]\label{def:NE}
	A collective vector of strategies $\xstar \in \X$ is a NE of \eqref{eq:problem} if, for all $i \in \mc I$,
	\begin{equation*}
		\Ji(\xistar, \agg(\xstar)) \leq \underset{\x_i \in \X_i}{\min} \, \Ji(\x_i,\tfrac{1}{N}\phii(\x_i) + \aggmi(\xstari)), 
	\end{equation*}
	where $\aggmi(x_{-i}^\star)\coloneqq \tfrac{1}{N} \sum_{j\in\mc I\setminus\{i\}} \phij(x_j^\star)$, and $x_{-i}^\star\coloneqq \col(x_1^\star,\ldots,x_{i-1}^\star,x_{i+1}^\star,\ldots,x_N^\star) \in \R^{n-n_i}$.
\end{definition} 
To demonstrate the relevance of this framework, we formulate a voltage support problem in smart grids as an instance of~\eqref{eq:problem}. We will then revisit this problem in \S \ref{sec:numerical_simulation} to illustrate the efficacy of the proposed scheme numerically.

\subsection{Case study: voltage support in smart grids}
\label{subsec:motivating_example}

Consider a set of $N$ electric vehicles (EVs) populating a distribution grid, in which
each agent controls both the active and reactive power injections of their own EV charger. 
Active power demand is exclusively associated to battery charging and is subject to the presence of a plugged-in vehicle. The charger is equipped with an inverter capable of full two-quadrant operation (see, e.g.,~\cite{Qatnight1}).
In this setting, every agent wishes to determine a day-ahead EV charging schedule, subject to time-varying cost of electricity and an additional fee applied by the distribution system operator (DSO) to encourage voltage support provision \cite{BolognaniEtAl2015TAC,DorflerEtAl2019ECC}. The latter is carried out by means of reactive power compensation: this is known to benefit the network efficiency, in terms of reduction of losses as well as curtailment of distributed generators \cite{DeakinEtAl2018PESGM,DEAKIN2020106770}.
Given a discrete time interval $\{1,\ldots,T\}$, we define each agent's strategy as $x_i\coloneqq \col(p_i,q_i)$, where $p_i,q_i\in\mathbb{R}^{T}$ are the active and reactive power injections, respectively.
The voltage at each of the $N_b \in \N$ buses is obtained through the linearized \emph{DistFlow} branch flow model as a function of nodal power injections \cite{BaranWu1989_TPD,FarivarEtAl2013_CDC}. In our problem setup, this is
\begin{displaymath}
	v = v_0 + \sum_{i=1}^N(\rho_i\otimes I_T)p_i + \sum_{i=1}^N(\xi_i\otimes I_T)q_i,
\end{displaymath}
where $v\in\mathbb{R}^{d}$, with $d=N_bT$, is the vector of bus voltages, 
$\rho_i, \xi_i\in\mathbb{R}^{N_b}$ are the columns of the resistance and reactance matrices, respectively, corresponding to the bus where agent $i$ is connected; finally, $v_0$ is a voltage offset due to non-manipulable generation/load. 
Then, by letting  $\Phi_i\coloneqq N [\rho_i \; \xi_i]\otimes I_T \in \R^{d \times 2T}$ for all $i$, we define
\begin{equation}\label{eq:distflow}
	\sigma(x) = \frac{1}{N}\sum_{i=1}^N \Phi_i x_i,
\end{equation}

This yields a game in the form of \eqref{eq:problem}, where
\begin{displaymath}
	J_i(x_i,\sigma(x)) = -\col(\pi,0_{T \times 1})^\top x_i + \|\sigma(x) - \bar{\sigma} \|^2_H + \| x_i \|^2_{\lwm}.
\end{displaymath}
Here, to incentivize voltage support provision, the deviation of the aggregate variable from a given reference voltage vector $\bar{\sigma}\in\mathbb{R}^{N_bT}$ is penalized via the positive definite matrix $H \in \R^{d \times d}$; this can be broadcast, e.g., by the distribution system operator \cite{ZhangEtAl2018_JSTSP}.
Energy prices are expressed by $\pi\in\mathbb{R}^{T}$, and $\lwm \in \R^{2T\times 2T}$ is a positive definite weight matrix. 

Finally, $X_i\subset\mathbb{R}^{2T}$ takes the form:
\begin{multline*}
	X_i \coloneqq  \left\{(p_i,q_i)\mid a_{i,\mathrm{ch}}^\top p_i = b_{i,\mathrm{ch}} \right\}
	\cap \left\{(p_i,q_i)\mid p_i\leq 0 \right\}\\
	\cap \left\{(p_i,q_i) \mid p_{i\tau}^2 + q_{i\tau}^2  \leq \bar{S}^2,\,  \tau = 1,\ldots, T\right\},
\end{multline*}
where the last two terms model the EV charger's apparent power limit $\bar{S} > 0$ (in kVA), and prevents any active power from being fed back to the grid; the first term imposes a total recharge target $b_{i,\mathrm{ch}} > 0$ (kWh) along the horizon, with any recharge event being possible compatibly with the plugged-in state of the vehicle (encoded by the vector $a_{i,\mathrm{ch}}\in\{0,1\}^{T}$).

\section{Nash equilibrium seeking algorithm}
\label{sec:unconstrained_algorithm}

\subsection{Algorithm description and main result}

In this section, we derive a novel equilibrium-seeking algorithm for problem~\eqref{eq:problem} working in a fully distributed fashion.
Indeed, we assume partial information, i.e., each agent $i$ is only aware of its own local information (e.g., the current local strategy or the local functions $\Ji$ and $\phii$) and exchanges tentative decisions with a subset of agents considered as neighbors, according to an underlying communication graph.

In particular, we assume the $N$ agents communicating over a directed graph $\cG = (\mc I, \cE)$, with $\cE \subset \mc I^2$, such that $i$ can receive information from agent $j$ only if the edge $(j,i)\in\cE$. 
The set of in-neighbors of $i$ is represented by $\cN_i \coloneqq \{j \in \mc I \mid (j,i) \in \cE\}$ (where also $i\in\cN_i$).
Graph $\cG$ is associated with a weighted adjacency matrix $\cW \in\R^{N\times N}$ whose entries satisfy $w_{ij} >0$ whenever $(j,i)\in \cE$ and $w_{ij} =0$ otherwise. 

We proceed with the derivation of our distributed algorithm. Let $\xit \in \R^{n_i}$ be the strategy chosen by agent $i$ at iteration $\iter \ge 0$, $\iter \in \N$.
We update each agent's strategy by performing a projected pseudo-gradient descent step, as
\begin{equation}\label{eq:desired_update}
    \xitp = \Pxi{\xit - \gamma \left( \nabla_{\x_i}\Ji(\xit, \agg(\xt))\right)},
\end{equation}
where $\gamma > 0$ plays the role of the gradient stepsize.
 We point out that the chain rule and the definition of $\agg(\xt)$ (cf.~\eqref{eq:sigma}) lead to $\nabla_{\x_i}\Ji(\xit, \agg(\xt)) = \nabla_1 J_i(\xit,\agg(\xt)) + \tfrac{\nabla \phii(\xit)}{N}\nabla_2 \Ji(\xit,\agg(\xt))$.
In our distributed setting, however, agent $i$ does not have access to the aggregate variable $\agg(\xt)$. 
 To compensate for this lack of information, we rely on the auxiliary variable $\zit \in \R^{d}$. 
 Thus, for all $i \in \mc I$, we introduce the operator $\tFi: \R^{n_i} \times \R^d \to \R^{n_i}$ defined as
 \begin{equation*}
   \tFi(\x_i,s) \coloneqq \nabla_1 J_i(\x_i,s) + \tfrac{\nabla \phii(x_i)}{N}\nabla_2 \Ji(x_i,s),
 \end{equation*}
 and, in accordance, we modify the update~\eqref{eq:desired_update} as 
\begin{equation}
    \xitp = \Pxi{\xit - \gamma \tFi\left(\xit,\phii(\xit) + \zit\right)},\label{eq:implementable_update}
\end{equation}
which meets the distributed requirements of our algorithm. 
Note that, in case $\zit\rightarrow -\phii(\xit)  + \agg(\xt)$,
then the implementable law~\eqref{eq:implementable_update} coincides with the one in~\eqref{eq:desired_update}. 
In particular, variable $\zit$ encodes the estimate of $\agg(\xit) - \phii(\xit)$. 
For this reason, we update each auxiliary variable $\zit$ according to the following causal version of the perturbed average consensus scheme:
\begin{equation}\label{eq:update_zit}
    \zitp = \sum_{j\in\cN_i}w_{ij}\zjt + \sum_{j\in\cN_i}w_{ij}(\phij(\xjt) - \phii(\xit)).
\end{equation}
This is indeed implementable in a distributed fashion, as it only requires communication with neighboring agents $j \in \cN_i$. 
If each $\xit$ is fixed, then it is possible to establish the global exponential stability of equilibrium $-\phii(\xit) + \sigma(\xt)$ for each dynamics \eqref{eq:update_zit}.
With an eye to singular perturbation, we then modify~\eqref{eq:implementable_update} by introducing a tuning parameter $\delta \in (0,1)$ and taking a convex combination step that allows for arbitrarily reducing the variations of each $\xit$ while preserving the invariance of each $\X_i$, thus obtaining:
\begin{equation*}
    \xitp = \xit + \delta\left(\Pxi{\xit - \gamma \tFi\left(\xit,\phii(\xit) + \zit\right)} - \xit\right).
\end{equation*}

We summarize our main update steps in Algorithm~\ref{algo:unconstrained}, which will be hereafter referred to as TRacking Aggregative Distributed Equilibrium Seeking (\algo/).
\begin{algorithm}[t!]
	\begin{algorithmic}
		\State \textbf{Initialization}: $\x_i^0 \in \R^{n_i}, \z_i^0 = 0$\\
		\smallskip
		\textbf{For} $t=1, 2, \dots$ \textbf{do} 
        \begin{subequations}\label{eq:local_update}
            \begin{align}
                \xitp\!&=\!\xit \!+\! \delta\left(\Pxi{\xit \!-\! \gamma\tFi\left(\xit,\phii(\xit) \!+\! \zit\right)} \!-\! \xit\right)\label{eq:x_update}
                \\
                \zitp &= \sum_{j\in\cN_i}w_{ij}\zjt + \sum_{j\in\cN_i}w_{ij}(\phij(\xjt) - \phii(\xit))\label{eq:z_local}
            \end{align}
        \end{subequations}
		\textbf{End}
	\end{algorithmic}
	\caption{\algo/ (Agent $i$)}
	\label{algo:unconstrained}
\end{algorithm}
Note that \algo/ requires the initialization $\z_i^0 = 0$ for all $i \in \mc I$: we will discuss in the sequel the interpretation of this specific choice. 
Algorithm~\ref{algo:unconstrained} appears similar to the scheme in~\cite{koshal2016distributed} where, however, a diminishing stepsize is employed and the convex combination step is missing.
As clarified later in the paper, such a combination step turns out to be crucial for using constant parameters $\delta$ and $\gamma$ and analyzing \algo/ under the lens of singular perturbation, ultimately yielding linear convergence to $\xstar$.
The local update~\eqref{eq:local_update} leads to the stacked vector form of \algo/, which reads as
\begin{subequations}\label{eq:global_update}
    \begin{align}
        \xtp &= \xt + \delta\!\left(\! P_{\X}\!\big[\xt \! - \! \gamma\tF\left(\xt,\phi(\xt) + \zt\right)\big] - \xt\!\right)\!\!,\label{eq:global_update_x}
        \\
        \ztp &= \Wd\zt + (\Wd - I_{Nd})\phi(\xt),\label{eq:global_update_z}
  \end{align}  
\end{subequations}
with $\Wd \coloneqq \cW \otimes I_d \in \R^{Nd}$, $\zt \coloneqq \col(\z_{1,\iter},\dots,\z_{N,\iter})$, $\phi(\xt) \coloneqq \col(\phi_1(\xt_1),\dots,\phi_N(\xt_N))$, $\tF(\xt,\phi(\xt) + \zt) \coloneqq \col(\tilde{F}_1(\x_{1}^{\iter},\phi_1(\x_{1}^{\iter}) + z_{1}^{\iter}), \dots,
\tilde{F}_N(\x_{N}^{\iter},\phi_N(\x_{N}^{\iter}) + z_{N}^{\iter}))$.

Before establishing the convergence properties of \algo/, we need to enforce the following assumptions.
\begin{assumption}[\textup{Feasible sets and cost functions}]\label{ass:local_feasible_set}
    For all $i \in \mc I$, we have that:
        \begin{itemize} 
            \item[(i)] The feasible set $\X_i$ is nonempty, closed, and convex;
            \item[(ii)] The function $J_i(\cdot,\phii(\cdot)/N + \sigmai(\xmi))$ is of class $\cC^1$, i.e., its derivative exists and is continuous, for all $\xmi$.             
        \end{itemize} 
    \end{assumption}
    A crucial element in this game-theoretic framework is the so-called \emph{pseudo-gradient mapping} $F: \R^n \to \R^n$:
    \begin{equation}\label{eq:F_definition}
        \hspace{-.2cm}F(\x) \coloneqq \col(\nabla_{\x_1} J_1(\x_1,\agg(\x)), \dots, \nabla_{\x_N} J_N(\x_N,\agg(\x))),
    \end{equation}
    for which we postulate the following conditions.
    \begin{assumption}[\textup{Strong monotonicity and Lipschitz continuity}]
        \label{ass:objective_function}
        $F$ is $\mu$-strongly monotone, i.e., there exists $\mu > 0$ such that, for any $\x, \y \in \R^{n}$,  $(F(x) - F(y))\T (\x - \y) \ge \mu\norm{\x - \y}^2.$
        Moreover, given any $\x_i, \x_i^\prime \in \R^{n_i}$ and $y, y^\prime \in \R^{n-n_i}$, for all $i \in \mc I$ we assume that 
        \begin{align*}
            \|\nabla_{\x_i}\Ji(\x_i,\phii(\x_i)/N + y) - & \nabla_{\x_i^\prime}\Ji(\x_i^\prime,\phii(\x_i^\prime)/N + y^\prime)\|
            \\
            & \leq \!\beta_1\!\norm{\col(\x_i,y) \!-\! \col(\x_i^\prime,y^\prime)}\!,
            \\
            \norm{\nabla_1\Ji(\x_i,y) \!-\! \nabla_1\Ji(\x_i^\prime,y^\prime)} & \leq \!\beta_1\!\norm{\col(\x_i,y) \!-\! \col(\x_i^\prime,y^\prime)}\!,
            \\
            \norm{\nabla_2\Ji(\x_i,y) \!-\! \nabla_2\Ji(\x_i^\prime,y^\prime)} & \leq \!\beta_2\!\norm{\col(\x_i,y) \!-\! \col(\x_i^\prime,y^\prime)}\!,
            \\
            \norm{\phii(\x_i) - \phii(\x_i^\prime)} &\leq \beta_3\norm{\x_i - \x_i^\prime}, 
        \end{align*} 
        hold true for positive scalars $\beta_1,\beta_2, \beta_3>0$.
    \end{assumption}
     
    Assumptions~\ref{ass:local_feasible_set} and~\ref{ass:objective_function} together guarantee the existence and uniqueness of a NE $\xstar$ for~\eqref{eq:problem}~\cite[Th.~2.3.3]{facchinei2003finite}.  
Further, for any $\gamma > 0$, it holds that \cite[Ch.~12]{facchinei2003finite}
\begin{align*}
    \xstar = \Px{\xstar - \gamma \F(\xstar)}.
\end{align*}
This, in turn, leads to $\xstar =\xstar + \delta(\Px{\xstar - \gamma \F(\xstar)} - \xstar)$ for any $\delta > 0$.
The following assumption characterizes the considered network graph.
\begin{assumption}[\textup{Network}]
    \label{ass:network}
    The graph $\cG$ is strongly connected, i.e., for every $(i,j) \in \mc I^2$ there exists a path of directed edges from $i$ to $j$, and the matrix $\cW$ is doubly stochastic, i.e., $\cW 1_N = 1_N$ and $1_N\T \cW = 1_N\T.$
\end{assumption}
We are now in the position of providing the convergence properties of \algo/ in computing the NE $\xstar$ of~\eqref{eq:problem}.
\begin{theorem}\label{th:convergence}
	Consider~\eqref{eq:global_update}.
	There exist constants $\bar{\gamma}, \bar{\delta}, a_1, a_2 > 0$ such that, for any $\delta \in (0,\bar{\delta})$ and $\gamma \in (0,\bar{\gamma})$ and $(\x^0,\z^0) \in \X \times \R^{Nd}$ such that $\oned\T \z^0 = 0$, it holds that
	\begin{align*}
		\norm{\xt - \xstar} \leq a_1 e^{-a_2\iter}.
	\end{align*}
\end{theorem}
The proof of Theorem~\ref{th:convergence} relies on a \emph{singular perturbation} analysis of~\eqref{eq:global_update} and it is provided in the next subsection.

\subsection{Proof of Theorem \ref{th:convergence}}
\label{sec:thm_proof}

The key intuition consists in the interpretation of~\eqref{eq:global_update} as a singularly perturbed system, i.e., an interconnected dynamics in the form 
\begin{subequations}\label{eq:interconnected_system_generic}
    \begin{align}
        \chitp &= \chit + \delta f(\chit,\wt)\label{eq:slow_system_generic}
        \\
        \wtp &= g(\wt,\chit,\delta),\label{eq:fast_system_generic}
    \end{align}
\end{subequations}
with $\chit \in \cD \subseteq \R^n$, $\wt \in \R^m$, $\delta > 0$, $f: \cD \to \R^n$, and $g: \cD \times \R^m \times \R \to \R^m$.
The subsystem~\eqref{eq:slow_system_generic} is the so-called slow dynamics and~\eqref{eq:fast_system_generic} is the so-called fast one having an equilibrium parametrized in the slow state $\chit$, namely, there exists $h: \cD \to \R^m$ such that 
\begin{align}
    0 = g(h(\chi),\chi,\delta),
\end{align}
for all $\chi \in \cD$ and $\delta > 0$.

Singular perturbations results (see, e.g.,~\cite[Th.~II.5]{carnevale2022tracking}) allow for guaranteeing the stability properties of interconnected systems of the form~\eqref{eq:interconnected_system_generic} by separately studying the so-called boundary layer system 
\begin{equation}\label{eq:boundary_layer_system_generic}
    \psitp = g(\psit + h(\chi),\chi,\delta) - h(\chi)
\end{equation}
with $ \psit \in \R^m$, and the so-called reduced system
\begin{equation}\label{eq:reduced_system_generic}
    \chitp = \chit + \delta  f(\chit,h(\chit)).
\end{equation}
We note that the boundary layer system~\eqref{eq:boundary_layer_system_generic} is obtained by arbitrarily fixing the slow state $\chit = \chi$ in the fast dynamics~\eqref{eq:fast_system_generic} and considering the error coordinate with respect to the parametrized equilibrium $h(\chi)$.
As for the reduced system~\eqref{eq:reduced_system_generic}, it is obtained by plugging $\wt = h(\chit)$ (i.e., fast state always at the parametrized equilibrium) into the slow dynamics~\eqref{eq:slow_system_generic}.
The rationale behind this approach is that, by decreasing $\delta$ (see~\eqref{eq:slow_system_generic}), we can arbitrarily reduce the variations of $\chit$ and, thus, also the approximations characterizing the auxiliary systems~\eqref{eq:boundary_layer_system_generic}-\eqref{eq:reduced_system_generic}.

We will analyze~\eqref{eq:global_update} scheme by identifying (a transformation of) the consensus-based part~\eqref{eq:global_update_z} as the fast dynamics and the gradient-based part~\eqref{eq:global_update_x} as the slow one.

The proof of Theorem~\ref{th:convergence} thus builds upon five steps:

\emph{1. Bringing \eqref{eq:global_update} in the form of \eqref{eq:interconnected_system_generic}:} We introduce the coordinates $\bz \in \R^d$ and $\pz \in \R^{(N-1)d}$ defined as follows:
    \begin{align}\label{eq:change_mean}
        \begin{bmatrix}
            \bz
            \\
            \pz
        \end{bmatrix} \coloneqq \begin{bmatrix}
            \tfrac{\oned\T}{N}\\
            \Rd\T
        \end{bmatrix}\z \implies \z = \oned\bz + \Rd\pz,
    \end{align}
    where $\Rd \in \R^{Nd \times (N-1)d}$ with $\norm{\Rd} = 1$ is such that
    \begin{align}
    \Rd \Rd\T = I_{Nd} - \tfrac{\oned\oned\T}{N}~\text{ and }~ \Rd\T\oned = 0. \label{eq:eq_Rd}
    \end{align}
    By combining the definition of $\bz$ in~\eqref{eq:change_mean}, the update~\eqref{eq:global_update_z}, the identities in~\eqref{eq:eq_Rd}, and the fact that $\oned\T \Wd = \oned\T$ and $\oned\T (\Wd - I) = 0$ (cf. Standing Assumption~\ref{ass:network}), we get
    \begin{align}
        \bztp = \bzt,\label{eq:barz_update}
    \end{align}
    which, in turn, leads to $\bztp = \bz^0 = 0$ for all $\iter \ge 0$ by the initialization $\oned\T z^0 = 0$.
    We are thus entitled to ignore the dynamics of $\bzt$ and, according to~\eqref{eq:change_mean}, we rewrite~\eqref{eq:global_update} as
    \begin{subequations}\label{eq:global_update_mean}
        \begin{align}
            \xtp\!&=\!\xt \!+\! \delta\big(P_X[\xt \!-\! \gamma \tF(\xt,\phi(\xt) \!+\! \Rd\pzt)] \!-\! \xt\big),\label{eq:x_update_mean}
            \\
            \pztp &= \Rd\T \Wd\Rd\pzt + \Rd\T(\Wd - I)\phi(\xt).\label{eq:z_update_mean}
      \end{align}  
    \end{subequations}
    For any $\iter \geq0$, the interconnected system~\eqref{eq:interconnected_system_generic} can thus be obtained from \eqref{eq:global_update_mean} by setting 
    \begin{align}\label{eq:assign}
        \begin{split}
            \chit &\coloneqq \xt, \quad w^t \coloneqq \pzt, 
            \\
            f(\chit,\wt) &\coloneqq \Px{\xt - \gamma \tF(\xt,\phi(\xt) + \Rd\pzt)} - \xt,
            \\
            g(\wt,\chit) &\coloneqq \Rd\T \Wd\Rd\wt + \Rd\T(\Wd - I)\phi(\xt). 
        \end{split}
    \end{align}
    In particular, we refer to the subsystem~\eqref{eq:x_update_mean} as the slow system, while we refer to~\eqref{eq:z_update_mean} as the fast one.

    \emph{2. Equilibrium function $h~$:} For any $\xt \in\R^n$, using~\eqref{eq:eq_Rd} and since $\cW$ is doubly stochastic (cf. Standing Assumption~\ref{ass:network}) notice that, for any $\xt = \x \in \R^n$, the point
    \begin{equation}\label{eq:man_definition}
        \pz = h(\x) \coloneqq -\Rd\T\phi(\x)
    \end{equation}
    constitutes an equilibrium of~\eqref{eq:z_update_mean}.
    Since $\Rd\T \Wd\Rd$ is Schur in view of Standing Assumption~\ref{ass:network}, we interpret~\eqref{eq:z_update_mean} as a strictly stable linear system with nonlinear input $\Rd\T(\Wd - I)\phi(\xt)$ parametrizing the equilibrium of the subsystem. 
    The role of $\gamma$ is to slow down the variation of $\xt$ so that the stability of $h(\xt)$ for~\eqref{eq:z_update_mean} is preserved.

    \emph{3. Boundary layer system:} 
    The so-called boundary layer system associated to~\eqref{eq:global_update_mean} can be constructed by fixing $\xt = x$ for all $\iter \geq0$, for some arbitrary $\x \in \mathbb{R}^n$ in \eqref{eq:z_update_mean}, and by rewriting it according to the error coordinates $\tzt := \pzt - h(\xt)$. 
    Using~\eqref{eq:eq_Rd}, the definition of $h$ (cf.~\eqref{eq:man_definition}), and the fact that $\cW$ is doubly stochastic (cf. Standing Assumption~\ref{ass:network}), we get
    \begin{equation}\label{eq:bl}
        \tztp = \Rd\T \Wd\Rd\tzt.
    \end{equation}
    Notice that the obtained system is in the form of \eqref{eq:boundary_layer_system_generic} with $\psi =\tzt$, and $g(\psi  +  h(\x),\x) - h(\x) = \Rd\T \Wd\Rd\tz$.
    The next lemma provides a Lyapunov function for \eqref{eq:bl}.
    \begin{lemma}[\textup{\cite[Lemma~III.4]{carnevale2022tracking}}]\label{lemma:bl}
        Consider system~\eqref{eq:bl}. 
        Then, there exists a continuous function $U :\R^{(N-1)d} \to \R$ such that 
        \begin{subequations}\label{eq:U_generic}
            \begin{align}
                &b_1 \norm{\tz}^2 \leq U(\tz) \leq b_2\norm{\tz}^2 \label{eq:U_first_bound_generic}
                \\
                &U(\Rd\T \Wd\Rd\tz) -  U(\tz)  \leq  -b_3\norm{\tz}^2\label{eq:U_minus_generic}
                \\
                &|U(\tz_1) \! - \! U(\tz_2)| \! \leq \! b_4 \! \norm{\tz_1 \! - \! \tz_2}\!\norm{\tz_1}
                \! + \! b_4\!\norm{\tz_1 \! - \! \tz_2}\!\norm{\tz_2}\!,\!
                \label{eq:U_bound_generic}
            \end{align}
        \end{subequations}
        for all $\tz,  \tz_1,  \tz_2 \in \R^m$, and some $b_1, b_2, b_3, b_4 > 0$.\oprocend
    \end{lemma}
    \emph{4. Reduced system:}
   The so-called reduced system can be obtained by plugging into \eqref{eq:x_update_mean} the fast state at its steady-state equilibrium, i.e., we consider $\pzt = h(\xt)$ for all $\iter \ge 0$. We thus have:
   \begin{equation}\label{eq:rs}
        \xtp = \xt + \delta(P_X[\xt - \gamma \tF(\xt,\phi(\xt) + \Rd h(\xt))] - \xt).\!
   \end{equation}
    Due to~\eqref{eq:eq_Rd} we have that $\tF(\xt,\phi(\xt) + \Rd h(\xt)) = \tF(\xt,\oned\agg(\xt)) = F(\xt)$, so~\eqref{eq:rs} is equivalent to 
    \begin{equation}\label{eq:rs_explicit}
        \xtp = \xt + \delta\left(\Px{\xt - \gamma F(\xt)} - \xt\right).
    \end{equation}
    The next lemma provides a Lyapunov function for \eqref{eq:rs}.
    \begin{lemma}[\textup{\cite[Lemma~III.5]{carnevale2022tracking}}]\label{lemma:rs}
        Consider system~\eqref{eq:rs}. Then, there exist a continuous function $W: \R^n \to \R$ and $\bar{\delta}_2, \bar{\gamma} > 0$ such that, for any $\delta \in (0,\bar{\delta}_2)$ and $\gamma \in (0,\bar{\gamma})$, it holds 
        \begin{subequations}\label{eq:W_generic}
            \begin{align}
                &c_1 \norm{\x - \xstar}^2 \leq W(\x) \leq c_2\norm{\x - \xstar}^2\label{eq:W_first_bound_generic}
                \\
                &W(\x + \delta  f(\x,h(\x)))  -  W(\x)  \leq  -\delta c_3\norm{\x - \xstar}^2\label{eq:W_minus_generic}
                \\
                &|W(\x_1)-W(\x_2)|\leq c_4\norm{\x_1-\x_2}\norm{\x_1 - \xstar}
                \notag\\
                &\hspace{3.1cm}
                + c_4\norm{\x_1-\x_2}\norm{\x_2 - \xstar},\label{eq:W_bound_generic}
            \end{align}
        \end{subequations}
        for all $\x, \x_1, \x_2, \x_3 \in \X$ and some $c_1, c_2, c_3, c_4 > 0$.\oprocend
    \end{lemma}

\emph{5. Lipschitz continuity of $f$, $g$ and $h$:}
As we will be invoking~\cite[Th.~II.5]{carnevale2022tracking}, we need to ensure that the Lipschitz continuity assumptions required by that theorem are satisfied. In particular, we require $f$ and $g$ in \eqref{eq:assign} to be Lipschitz continuous w.r.t. both arguments and $h$ in \eqref{eq:man_definition} to be Lipschitz continuous w.r.t. $x$.
Since $g$ and $h$ are linear functions, then their Lipschitz property is trivially guaranteed.
As for $f$, the Lipschitz continuity follows by the one of $\nabla J_i$ (cf. Standing Assumption~\ref{ass:objective_function}).
By combining Lemma~\ref{lemma:bl} and \ref{lemma:rs} with the Lipschitz conditions expressed above,~\cite[Th.~II.5]{carnevale2022tracking} can therefore be applied. 
Thus, there exists $\bar{\delta} \in (0,\bar{\delta}_2)$ so that $(\xstar,h(\xstar))$ is an exponentially stable equilibrium for~\eqref{eq:global_update_mean}.

\section{Voltage support example revisited}
\label{sec:numerical_simulation}

\begin{table}[tb]
	\caption{Simulation parameters}
	\label{tab:sim_val}
	\centering
	\begin{tabular}{llll}
		\toprule
		Param.  & Unit & Description   & Value \\
		\midrule
		$T$ & h & Time interval & $24$\\
		$N$ &  & Number of EVs & $321$\\
		$N_b$ &  & Number of buses & $94$\\
		$\rho_i, \xi_i$ & p.u. & Impedance matrix & from \cite{PIRES2012313} \\
		$\pi$ & \EUR/kWh & Day-ahead prices  & from \cite{entsoe} \\
		$H$ & & Weight matrix & $I_{d}$\\
		$\lwm$ & & Local weight matrix & $\mathrm{diag}(1,10) \otimes I_T$\\
		$a_{i,\mathrm{ch}}$ &  & Plugged-in state & from \cite{My_EV_ave}\\
		$b_{i,\mathrm{ch}}$ & kWh & Recharge target & $\sim \mc U(0,40)$\\
		$\bar S$ & kVA & Inverter capacity & $7$\\
		$\eta$ & &  Graph edge inclusion prob. & $0.7$\\
		$\gamma$ & & \algo/ step size & $0.01$\\
		$\delta$ & & \algo/ parameter & $0.5$\\
		\bottomrule
	\end{tabular}
\end{table}

\begin{figure}[tb]
	\centering
	\includegraphics[width=\columnwidth, trim=0 1mm 5mm 3mm,clip]{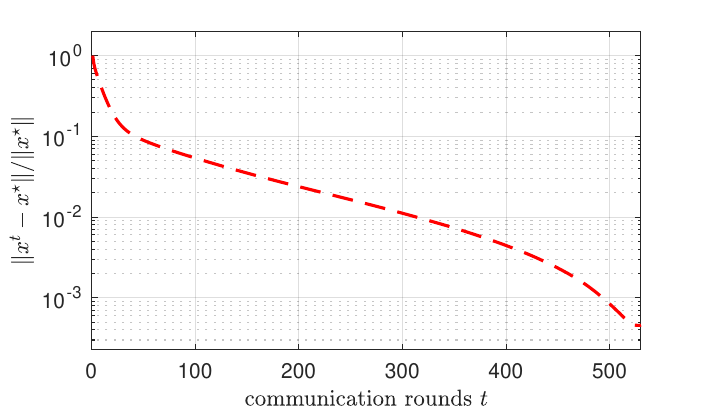}
	\caption{Convergence of the TRADES algorithm to the NE $\xstar\in\prod_{i=1}^{321} X_i$.
}
	\label{fig:conv_plot}
\end{figure}

\begin{figure}[tb]
	\centering
	\includegraphics[width=\columnwidth,trim=0 1mm 5mm 5mm,clip]{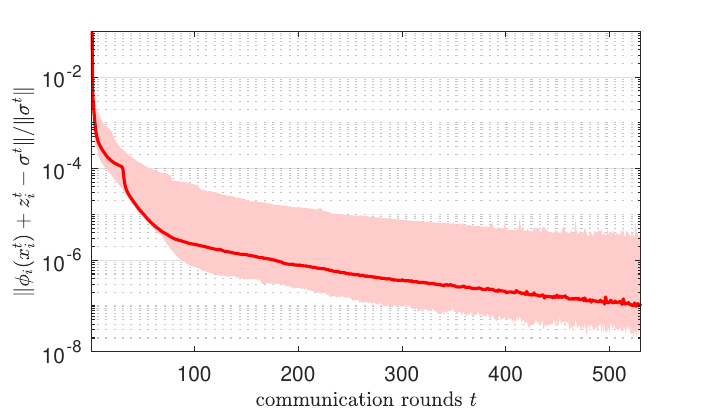}
	\caption{Error in the distributed estimation $\phi_i(x_i^t)+z_i^t$ of the bus voltage vector $\sigma^t\in\mathbb{R}^{N_bT}$, across iterates $t$. The shaded area refers to all $N$ agents, a median case is shown by the solid red line.}
	\label{fig:sigma_err}
\end{figure}

\begin{figure}[tb]
	\centering
	\includegraphics[width=\columnwidth,trim=3mm 1mm 0 4mm,clip]{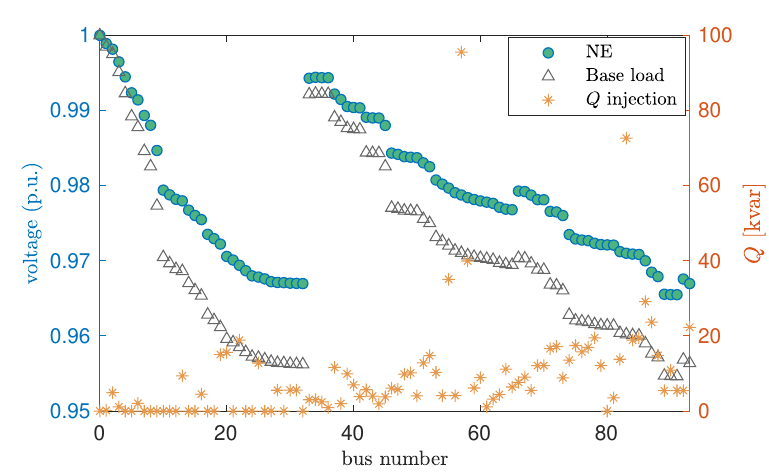}
	\caption{Variation of bus voltages due to the additional EV loads in the network (green circles), corresponding to the time interval where most EV charge occurs (12--1am). The resulting voltage distribution improves over the base load case (triangle markers), following reactive power injection from the smart EV chargers. The aggregate reactive injections at each bus are represented by asterisks; these are consistent with the considered network topology, where buses 57 and 83 are significantly loaded, thus requiring additional effort to compensate for voltage drop.}
	\label{fig:bus_volt_Q}
\end{figure}

We now revisit the example outlined in \S \ref{subsec:motivating_example}, and investigate the performance of \algo/ on an instance with $N=321$ agents. 
The numerical results shown in Figs.~\ref{fig:conv_plot}--\ref{fig:bus_volt_Q} are relative to a 94-bus radial network reflecting a real distribution system in Portugal~\cite{PIRES2012313}; parameters are listed in Tab.~\ref{tab:sim_val}. Agents are randomly assigned to buses proportionally to the baseline load (given in \cite{PIRES2012313}); they communicate over a directed, connected \er/ graph with edge inclusion probability $\eta$.   
Deviation from the nominal voltage (1 p.u.) at all buses was penalized by setting $\bar{\sigma} = \mathbf{1}_{N_b}-v_0$ (p.u.). 

Figure \ref{fig:conv_plot} demonstrates convergence of \algo/ to the NE $\xstar\in\prod_{i=1}^{321} X_i$. Figure \ref{fig:sigma_err}, instead, compares the actual bus voltage vector $\sigma^t\in\mathbb{R}^{N_bT}$ with its distributed estimate $\phi_i(x_i^t)+z_i^t$, across the algorithm iterates. 
The shaded area refers to their deviation across all agents, while the median case is shown by the solid red line.
We further stress here that the handling of \emph{generalized aggregators}~\cite{Jensen2010} as the one in~\eqref{eq:distflow} -- different from the arithmetic average standard in the cognate literature -- is another distinct feature of our method.

An excerpt of the resulting voltage profile, relative to the time interval 12-1 am, is illustrated in Fig.~\ref{fig:bus_volt_Q} with the corresponding nodal reactive power compensation, provided from the EV chargers at the obtained NE.
This yields an improvement of the voltage distribution over the base load case (triangle markers). 
The aggregate reactive injections at each bus are represented by asterisks; these are consistent with the considered network topology, where buses 57 and 83 are significantly loaded, thus requiring additional effort to compensate for voltage drop.


\section{Conclusion}
\label{sec:conclusions}

We have proposed a novel fully-distributed algorithm for NE seeking in aggregative games over networks. 
The convergence analysis of the underlying scheme has been developed under a singular perturbation lens. 
Specifically, slow and fast dynamics were identified and separately investigated to demonstrate the linear convergence of their system interconnection to an NE of the locally constrained games.
Current work focuses on a primal-dual scheme able to concurrently deal with coupling constraints (see, e.g.,~\cite{carnevale2022tracking}) and local ones while preserving linear convergence.

\section{Acknowledgments}
This work is funded in part by the European Union - NextGenerationEU under the National Recovery and Resilience Plan (PNRR) - Mission 4 Education and research - Component 2 From research to business - Investment 1.1 Notice Prin 2022 - DD n.~104 2/2/2022, from title ECODREAM Energy COmmunity management: DistRibutEd AlgorithMs and toolboxes for efficient and sustainable operations, proposal code 202228CTKY002 - CUP J53D23000560006.
F.~Fele acknowledges support from grant RYC2021-033960-I funded by MICIU/AEI/ 10.13039/501100011033 and European Union NextGenerationEU/PRTR, and grant PID2022-142946NA-I00 funded by MICIU/AEI/ 10.13039/501100011033 and ERDF/EU.

\bibliographystyle{IEEEtran} 
\bibliography{games_biblio}  

\end{document}